\documentclass[pra,amsmath,twocolumn,showpacs]{revtex4}
\usepackage{graphicx}
\usepackage{amsmath}
\usepackage{epigraph}
\begin{document}
\def\la{{\langle}}
\def\ra{{\rangle}}
\def\vep{{\varepsilon}}
\newcommand{\beq}{\begin{equation}}
\newcommand{\eeq}{\end{equation}}
\newcommand{\beqa}{\begin{eqnarray}}
\newcommand{\eeqa}{\end{eqnarray}}
\newcommand{\q}{\quad}
\newcommand{\tunn}{\text{tunn}}
\newcommand{\refl}{\text{refl}}
\newcommand{\all}{\text{all}}
\newcommand{\ion}{\text{ion}}
\newcommand{\bound}{\text{bound}}
\newcommand{\free}{\text{free}}
\newcommand{\lc}{\curly{l}}
\newcommand{\A}{\hat{A}}
\newcommand{\Pii}{\hat{\Pi}}
\newcommand{\s}{\hat{S}}
\newcommand{\x}{x_{cl}}
\newcommand{\si}{\hat{\sigma}}
\newcommand{\Pp}{\hat{\Pi}}
\newcommand{\AC}{{\it AC }}
\newcommand{\La}{{\lambda }}
\newcommand{\psii}{|\psi_I\ra}
\newcommand{\psif}{|\psi_F\ra}
\newcommand{\n}{\\ \nonumber}
\newcommand{\nn}{\q\q\q\q\q\q\q\q\q\\ \nonumber}
\q\q\q\q\q\q\q\q\q
\newcommand{\om}{\omega}
\newcommand{\U}{\hat{U}}
\newcommand{\up}{\hat{U}_{part}}
\newcommand{\mf}{m_f^{\alpha}}
\newcommand{\e}{\epsilon}
\newcommand{\Om}{\Omega}
\newcommand{\Tau}{\mathcal{T}_{SWP}}
\newcommand{\Ttu}{\tau_{in/out}}
\newcommand{\br}{\overline}
\newcommand{\cn}[1]{#1_{\hbox{\scriptsize{con}}}}
\newcommand{\sy}[1]{#1_{\hbox{\scriptsize{sys}}}}
\newcommand{\pd }{Pad\'{e} }
\newcommand{\PAD }{Pad\'{e}\q}
\newcommand{\PP }{\hat{\Pi}}
\newcommand{\su}{{\phi \gets \psi}}
\newcommand{\get }{\leftarrow}
\newcommand{\Eq}{equation }
\newcommand{\f}{\ref }
\newcommand{\T}{\text{T}_\Om}
\newcommand{\Tf}{\text{T}}
\newcommand{{\ttau}}{\overline{\tau_\Om} }
\newcommand{{\tttu}}{\overline{\tau_{[0,d]}} }
\newcommand{\h}{\hat{H}}
\newcommand{\N}{\mathfrak{N} }
\newcommand{\I}{\text{Im } }
\title{An even simpler understanding of quantum weak values}
\author {D. Sokolovski$^{1,3}$ and E. Akhmatskaya$^{2,3}$ }
\affiliation{$^1$ Departmento de Qu\'imica-F\'isica, Universidad del Pa\' is Vasco, UPV/EHU, Leioa, Spain}
\affiliation {$^2$ Basque Center for Applied Mathematics (BCAM),\\ Alameda de Mazarredo, 14 48009 Bilbao, Bizkaia, Spain}
\affiliation{$^3$ IKERBASQUE, Basque Foundation for Science, Maria Diaz de Haro 3, 48013, Bilbao, Spain}

\begin{abstract}
\noindent
ABSTRACT: 
\newline 
{We explain the properties and clarify the meaning of quantum weak values using only the basic notions of elementary quantum mechanics.}  

\date{\today}
\end{abstract}
\maketitle
\epigraph{...And look not for answers where no answers can be found.}{Bob Dylan}

\vskip0.5cm
In a recent publication \cite{Qin} Qin and co-authors sought to provide a simplified understanding of the physics of the so-called weak measurements (for a recent review see \cite{Rev1}). They formulated their discussion in the framework of the quantum Bayesian approach \cite{QBayes}, and followed other authors \cite{COMM1}, \cite{COMM2} in asserting that "anomalous" weak values (WV) may not occur in a purely classical context. One may wonder whether a yet more straightforward explanation of these properties could be obtained directly from the basic principles of quantum theory. In the following, we will provide such an explanation.
\newline
In quantum mechanics, e.g., in its field and many-body versions, the quantity of interest is often the probability $P^{\phi \gets \psi}$ for the system to start in an initial state $\psi$ and end up, after some time, in a final state $\phi$. The resulting probabilities obey all the rules of the classical probability theory, but the quantum nature of the problem dictates that in order to evaluate 
$P^{\phi \gets \psi}$, one must first obtain a complex valued transition {\it probability amplitude} $A^{\phi \gets \psi}$ \cite{FeynL}, so that
 \begin{eqnarray}\label{a1}
P^{\phi \gets \psi}=|A^{\phi \gets \psi}|^2.
\end{eqnarray}
Typically, an amplitude can be decomposed into various sub-amplitudes, corresponding to elementary processes, which all lead to the same outcome $\phi$, 
 \begin{eqnarray}\label{a2}
A^{\phi \gets \psi}=\sum_{n}A_n^{\phi \gets \psi}.
\end{eqnarray}
For example, for a system of interacting particles, $A_n^{\phi \gets \psi}$ could correspond to Feynman diagrams describing 
various scattering scenarios \cite{Matt}. The scenarios are "virtual", in the sense that only the probability amplitudes, and not the probabilities, can be ascribed to them individually. Together, virtual scenarios form a "real" pathway, connecting $\psi$ with $\phi$, which the system will be seen as taking with the probability (\ref{a1}), if the experiment is repeated many times. 

A simple illustration of the above is the Young's double slit experiment, sketched in Fig.1a.
An electron starts at some location $(x,y)$, and ends up in a final position $(x',y')$, which it can reach through two 
holes made in the screen. There are two virtual pathways, passing through the holes $1$ and $2$, with the probability amplitudes $A^{(x',y') \gets (x,y)}_1$ and $A^{(x',y') \gets (x,y)}_2$, respectively. A well known feature of quantum description is the impossibility do decide which of the two routes was actually taken. Any attempt to accurately determine it, destroys the  interference pattern, by changing the probability $P^{(x',y') \gets (x,y)}$ from $|A^{(x',y') \gets (x,y)}_1+A^{(x',y') \gets (x,y)}_2|^2$ to $|A^{(x',y') \gets (x,y)}_1|^2+|A^{(x',y') \gets (x,y)}_2|^2$. If no such attempt is made, "one may {\it not} say that an electron
 goes either through hole $1$ or hole $2$" \cite{FeynL}. The two virtual routes together form for the electron a single 
 real pathway from $(x,y)$ to $(x',y')$. This is the {\it uncertainty principle} \cite{FeynL}.

A further simplification of the double slit experiment, which  brings us closer to issue of weak values , is shown in Fig. 1b. Let a system, consisting of spin $1/2$, start in a state $|\psi\ra$ at $t=0$, evolve with a Hamiltonian $\h$ until $t=T$, and then be observed in the final state $|\phi\ra$.  Choosing an arbitrary basis $\{|i\ra\}$, $\la i| j\ra=\delta_{i,j}$, $i=1,2$, and inserting the unity $\sum_{i=1}^2 |i\ra\la i|$ at $t=T/2$, we can write the transition amplitude 
$A^{\phi \gets \psi}\equiv\la \phi|\exp(-i\h T)|\psi\ra$  as 
 \begin{eqnarray}\label{a3}
A^{\phi \gets \psi}=A_1^{\phi \gets \psi}+A_2^{\phi \gets \psi},
\end{eqnarray}
where ( $i=1,2$).
 \begin{eqnarray}\label{a3}
A_i^{\phi \gets \psi}=\la \phi|\exp(-i\h T/2)|i\ra\la i|\exp(-i\h T/2)|\psi\ra.\q
\end{eqnarray}
Unless one of the states $|i\ra$ coincides with $|\psi\ra$ or $|\phi\ra$, we have an analogue of the double slit experiment, with  $|i\ra$'s playing the role of the two holes, 
and $|\phi\ra$ representing the final position of the electron.
Our intention is to see whether the first "hole" was chosen by the system.
To obtain a yes/no answer,
we couple the spin to a von Neumann pointer at $t=T/2$,
using the interaction Hamiltonian, 
 \begin{eqnarray}\label{a3a}
\h_{int}=-i\partial_{f}\Pii_1\delta(t-T/2),
\end{eqnarray}
where $\Pii_1=|1\ra\la 1|$ is the projector on the state $|1\ra$ and $f$ stands for the pointer's position. 
\newline
Before the meter fires, the pointer's state
is $|G\ra$, and $G(f) \equiv \la f|G\ra$ is a real differentiable function (e.g., a Gaussian) with a zero mean, 
$\la G|f|G|\ra=0$. It peaks around $f=0$, 
and has a width $\Delta f$. At $t=T$, after a successful post-selection in $|\phi\ra$ the entangled state of the composite system,
$|\Phi(T)\ra$, is
 \begin{eqnarray}\label{a4}
\la f |\Phi(T)\ra=[G(f-1)A_1^{\phi \gets \psi} + G(f)A_2^{\phi \gets \psi}]|\phi\ra.
\end{eqnarray}
\begin{figure}
	\centering
		\includegraphics[width=6cm,height=10 cm]{{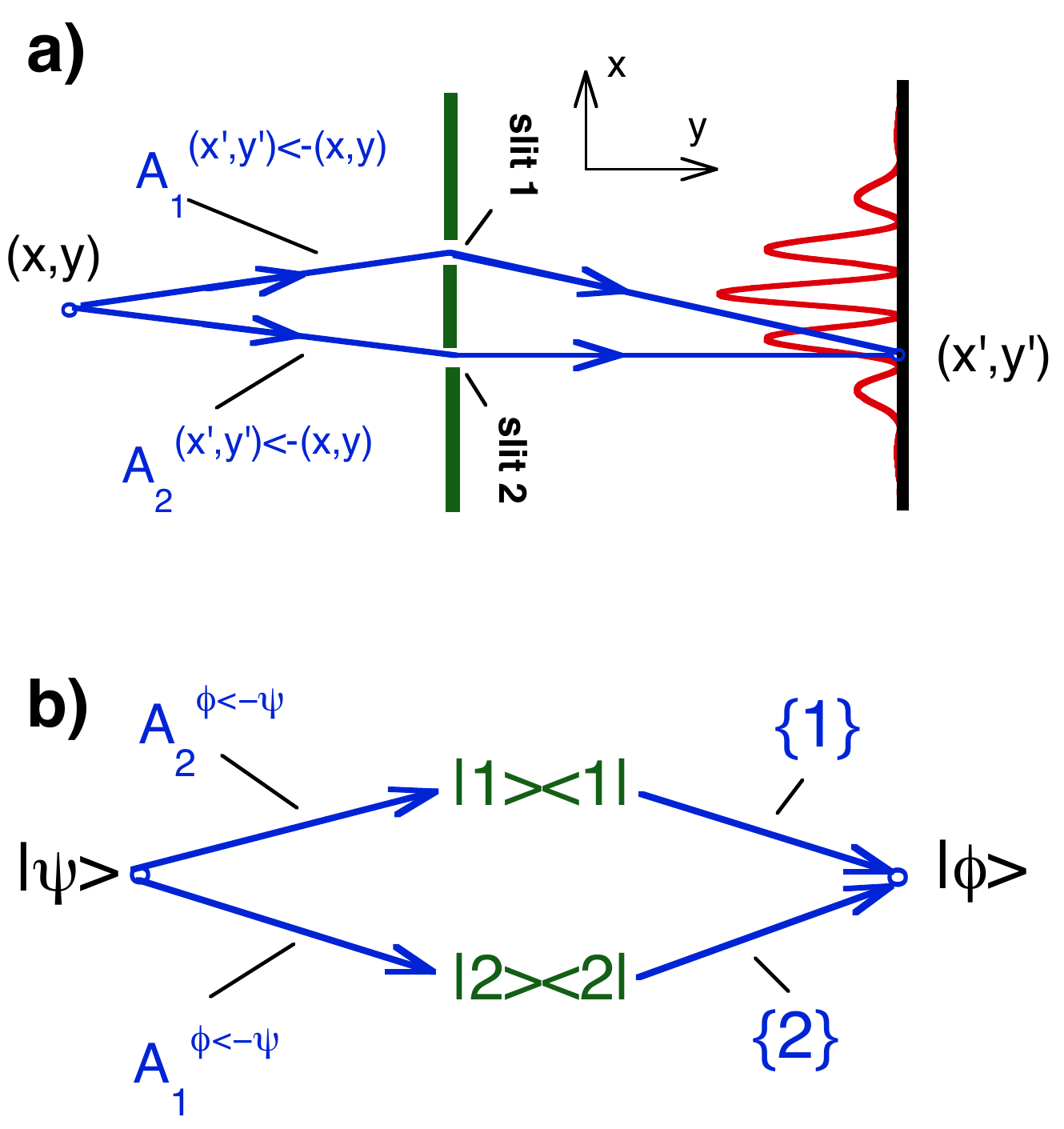}}
\caption{
\textbf{(a)} Young's two-slit experiment, where the initial and final position of the electron are connected by two virtual pathways, each passing through one of the slits; \textbf{(b)} in a simplified version of the experiment, the initial, $|\psi\ra$, and final, $|\phi\ra$ of a spin $1/2$ are connected by two virtual pathways, $\{1\} =|\phi\ra \gets |1\ra \gets |\psi\ra$ and $\{2\} =|\phi\ra \gets |2\ra \gets |\psi\ra$. The corresponding probability amplitudes are given by \Eq(\ref{a3}).} 
\label{fig:3}
\end{figure}
What we see depends on how hard we look, and this is determined by the choice of $\Delta f$.
For $\Delta f << 1$, only one of the two terms in (\ref{a4}) can have a non-zero value, and in  each run of the experiment  the pointer will be shifted by either $0$, or $1$. If the experiment is repeated many times, $N>>1$, for the mean pointer shift
 we find
 \begin{eqnarray}\label{a5a}
 \la f \ra^\su =\frac{\int \la\Phi(T)|f\ra f\la f| \Phi(T)\ra df}{\la\Phi(T)| \Phi(T)\ra}=
 \n
\frac{|A_1^{\phi \gets \psi}|^2}{|A_1^{\phi \gets \psi}|^2+|A_2^{\phi \gets \psi}|^2}.\q
\end{eqnarray}
\newline
We can also look at the situation from a purely classical point of view.
Accurate intermediate measurements destroy all interference between 
the virtual routes, turning them into two exclusive (real) alternatives \cite{FeynL}, and we can see which of the two is actually taken.
Out of $N$ trials, the pointer will read $1$ in $N_1$ cases, and $0$, in $N_2$ cases.
Thus the routes will be seen as travelled with the probabilities $\om_i$, $i=1,2$
 \begin{eqnarray}\label{a5}
\om_i^\su = lim_{N\to \infty} \frac{N_i}{N} =
\frac{|A_i^{\phi \gets \psi}|^2}{|A_1^{\phi \gets \psi}|^2+|A_2^{\phi \gets \psi}|^2}.
\end{eqnarray}
With this we can determine the mean value of the projector $\Pii_1$ simply by writing down $1$ 
whenever the spin is seen to pass via the state $|1\ra$, and  $0$ when it passes 
via the state $|2\ra$, add up the results, and divide by the number of trails $N$, 
 \begin{eqnarray}\label{a6}
\la \Pii_1\ra ^\su=\frac{1\times N_1+ 0\times N_2}{N}=\om_1^\su=\la f \ra^\su.
\end{eqnarray}
\newline
We can number the routes as $1$ and $2$, and ask the "which route?" question,
 by measuring instead of $\Pii_1$ the "route number operator"
 \begin{eqnarray}\label{a4a}
\hat{n}=|1\ra n_1 \la 1|+|2\ra n_2 \la 2|,\q n_i=i, \q i=1,2,
\end{eqnarray}
so that our accurate meter reads $1$ or $2$, depending on whether the system passes through the states $|1\ra$ or $|2\ra$.
By linearity, the mean value 
of any operator with the eigenvalues $B_i$, $\hat{B} =\sum_{i=1}^2 |i\ra B_i\la i|$
must be given by 
 \begin{eqnarray}\label{a6}
\la \hat{B}\ra ^\su=\om_1^\su B_1+\om_2^\su B_2,
\end{eqnarray}
and the mean route number is $\la \hat{n}\ra ^\su = \om_1^\su +2\om_2^\su$.  We note that $1\le \la \hat{n}\ra ^\su \le 2$,
and, from the position of $\la \hat{n}\ra ^\su$ inside the interval, it is possible to decide which of the two routes is travelled more often.
\newline
So far, there has been little "quantum" in our attempt to answer the "which way?" question. Out of the original quantum system, we have "manufactured"  
a simple classical system,  capable of reaching its final state by taking one of the two available paths at random (see Fig.2a).
The measured operator is replaced by a functional on the paths, whose mean value is obtained by recording 
$B_1$ or $B_2$, depending on the path taken, summing all values, and dividing the result by the number of trials $N$.
\newline
The quantum nature of the problem comes to light if one tries to answer the "which way? question with the interference between the routes intact.
We may try to use the same meter to measure $\hat{n}$ in equation (\ref{a4a}), but this time making sure that no new "real" (as opposed to "virtual")  routes are created for the system, and the transition is perturbed as little as possible.
One way to achieve this is to make the meter highly inaccurate, by choosing $G(f)$ so broad, that $G(f-1)\approx G(f-2) \approx G(f)$, and  \Eq(\ref{a4}) becomes $\la f |\Phi(T)\ra\approx G(f)A^{\phi \gets \psi}$. As a result,
the pointer's readings become equally spread over the whole real axis. This is what we should expect from the uncertainty principle, which suggests that number of the route taken by the spin, like the number of the slit taken in Fig.1a, remains {\it indeterminate}, provided the routes interfere. 
Indeed, according to Feynman \cite{FeynL}, this "which way?" question cannot be answered, 
and our experiment gives the only answer possible under the circumstances, which must be {\it "anything at all"}.
\newline
The mean pointer reading is, however, uniquely defined for any choice of the initial and final states. In an accurate measurement of the projector $\Pii_1$, $ \la f \ra^\su$ in \Eq(\ref{a5a})  coincided with the probability, with which the first path is travelled. What would it be with the interference intact? Using the definition (\ref{a5a}), expanding now very broad $G(f-1)$ in a Tailor series around $f=1$, and retaining only the leading terms, we find 
 \begin{eqnarray}\label{a7}
\la f\ra^\su\approx \text{Re}\left \{\frac{A_1^{\phi \gets \psi}}{A_1^{\phi \gets \psi}+A_2^{\phi \gets \psi}}\right \}\equiv \text{Re}\{\alpha_1^{\phi \gets \psi}\},
\end{eqnarray} 
where $\alpha_i^{\phi \gets \psi}=A_i^{\phi \gets \psi}/\sum_{j=1}^2 A_j^{\phi \gets \psi}$ are the probability amplitudes, renormalised to sum to unity.
Thus, with the interference present, the probability $\om_1$, with which the real first route is travelled in \Eq(\ref{a5a}), is substituted with 
the real part of the (relative) probability amplitude for the first virtual route. It is easy to check that if the projector $\Pii_1$ is replaced by an arbitrary operator $\hat{B}$, the mean shift of the pointer in the limit $\Delta f\to \infty$ is given by the real part of a sum of  $\alpha_i^{\phi \gets \psi}$, weighted by the eigenvalues of $\hat{B}$,
 \begin{eqnarray}\label{a9}
\la f \ra^\su \approx \text{Re} \left \{ \sum_{i=1}^2 B_i \alpha_i^{\phi \gets \psi}\right \},
\end{eqnarray}
so that for the route number $\hat{n}$ in  \Eq (\ref{a4a}) we find $\la f \ra^\su=\alpha_1^{\phi \gets \psi}+2\alpha_2^{\phi \gets \psi}$.
\begin{figure}
	\centering
		\includegraphics[width=6cm,height=4cm]{{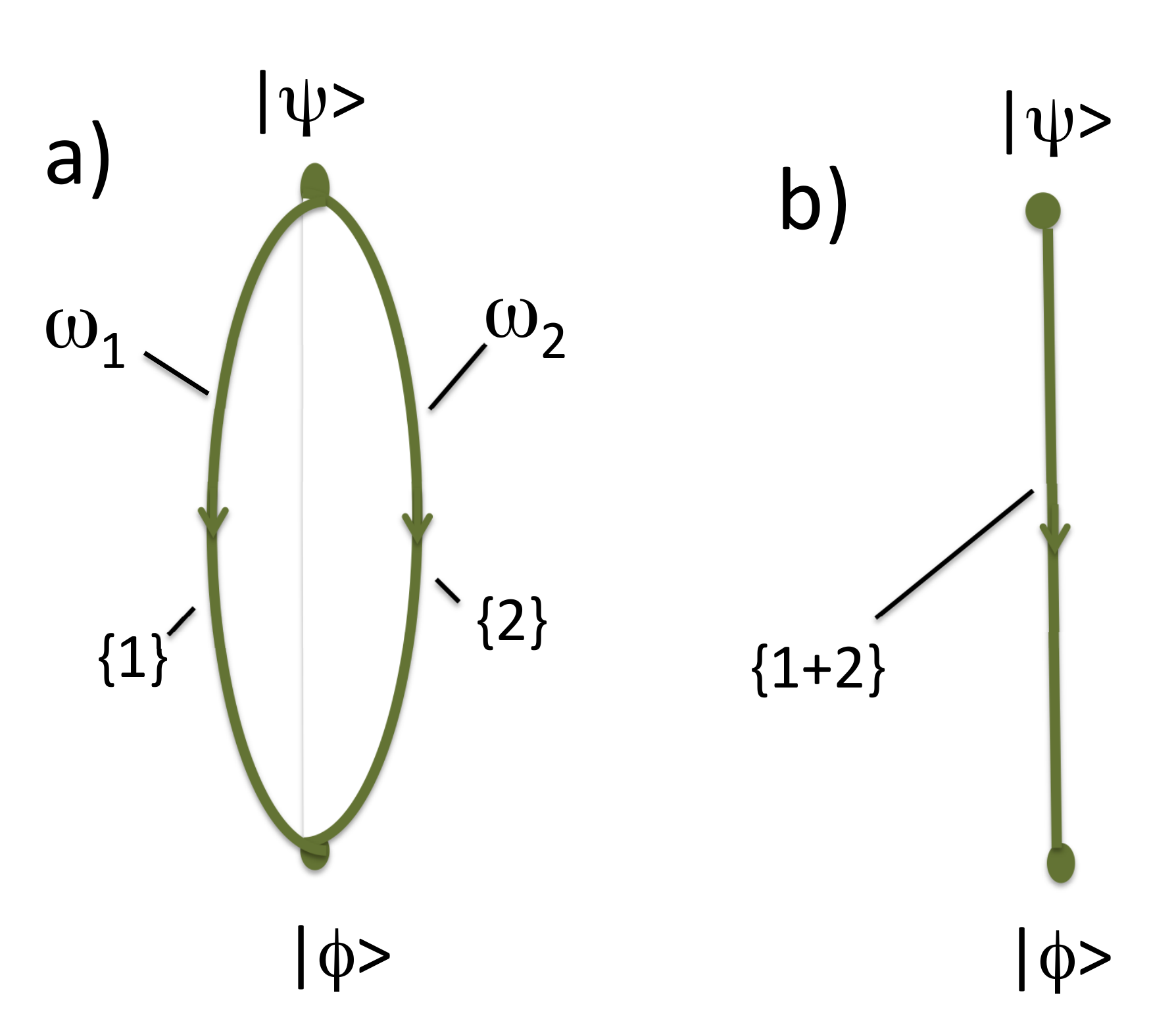}}
\caption{
\textbf{(a)}  An accurate measurement of the projector $\Pii_1$ creates two alternative pathways, $\{1\}$ and $\{2\}$, which the spin can be seen to take 
with the probabilities $\om_i^\su$ in \Eq(\ref{a5}); \textbf{(b)}  in an inaccurate (weak) measurement there is a single real pathway, 
arising from the interference between the virtual paths $\{1\}$ and $\{2\}$. 
}
\label{fig:3}
\end{figure}

We can now formulate a simple principle for intermediate measurements made on pre- and post-selected quantum system.
In  an accurate measurement, the mean shift of the pointer is given by a sum of the {\it probabilities} on the real paths the measurement creates, weighted by the eigenvalues of the measured operator. In a highly inaccurate measurement, this mean shift is expressed in terms of  the linear combinations of the corresponding {\it probability amplitudes} \cite{FOOT}. These amplitudes are, indeed, measured experimentally (see, for example, \cite{Lund}).
Note that we are no closer to resolving the original "which way?" conundrum. The uncertainty principle is still in place, and knowing only the amplitudes does not allow us to say which of the two ways was actually taken \cite{FeynL}.
\newline
It may, therefore, seem contradictory, that a {\it definite} result is obtained for each choice of the initial and final states, 
whereas the uncertainty principle appears to forbid obtaining any useful information regarding interfering alternatives. 
One sees, however, that the principle holds, since, in an inaccurate measurement of any operator  $\hat{B}$, it is possible to obtain {\it any} pointer shift, by selecting different initial and final states for the system. Indeed, choosing some $|\psi\ra =(a_1|1\ra + a_2|2\ra)/(a_1^2+a_2^2)^{1/2}$, with $a_i\ne 0$, 
and $|\phi\ra =(b_1|1\ra + b_2|2\ra)/(b_1^2+b_2^2)^{1/2}$, for $\h=0$ \cite{FOOT2} we can write the relative amplitudes  as 
$\alpha_i^{\phi \gets \psi}=\eta_i/(\eta_1+\eta_2)$, where (a star denotes complex conjugation) $\eta_i=b_i^*a_i$. Next we note that the equation
 \begin{eqnarray}\label{a10}
\frac{B_1\eta_1+B_2\eta_2}{\eta_1+\eta_2}=Z
\end{eqnarray}
always has a solution, so that for any given $|\psi\ra$ it is always possible to find a $|\phi\ra$, such that in \Eq(\ref{a9}) the complex valued quantity in the curly brackets takes any complex value $Z$. Thus, with all final states considered, the mean shift of an inaccurate pointer can again be {\it anything at all}, for any choice of the operator $\hat{B}$. This, in turn, means that the significance of a result, obtained for a given choice of  $|\psi\ra$ and $|\phi\ra$, is limited strictly to the particular condition under which the experiment is made. For example, ensuring a large value of $\text{Re}Z$, might help to amplify the deflection of  an electron beam \cite{Duck} or optical \cite{Metro} beam, but would provide no insight into the nature of the electron or the photon, beyond what is already known.
\newline
The mechanism, which allows $Z$ in  \Eq(\ref{a10}) to take an arbitrary value is simple. 
The l.h.s. of (\ref{a10}) has the form of an average, computed with a distribution $\eta_i$, which can take complex values. 
What is more important, its real and imaginary parts do not have to have definite signs. For example, by choosing $|\psi\ra$ and $|\phi\ra$ to be nearly orthogonal, $\eta_1 \approx -\eta_2$,
one can make the denominator of the ratio in (\ref{a10}) very small,
while its numerator remains finite, in which case the mean shift of an inaccurate pointer will be very large.
Note that this could never happen in an accurate measurement, since both $\om_i^\su$ in \Eq(\ref{a6}) are non-negative, 
and the mean shift of an accurate meter cannot exceed the larger of the eigenvalues $B_i$, nor be smaller than the smaller one.
By the same token, such anomalously large values cannot occur in purely classical theories, operating only with non-negative probabilities, contrary to the suggestion made in the much criticised (see \cite{Qin},\cite{COMM1},\cite{COMM2}, and Refs. therein)   work by Ferrie and Combes \cite{FC}.

In the title we have promised to provide a simple understanding of quantum weak values, a task we have avoided mentioning so far. Above we have demonstrated that a response of quantum system to probe by a particular weak interaction is formulated in terms of the corresponding probability amplitudes. This could be anticipated from a textbook on perturbation theory. 
We have also checked that the results are in full agreement with the uncertainty principle, and are, to a large degree, dictated by it. Next we try to describe these results in terms of the "weak values", as they were introduced in \cite{Ah1}, and find the origin of the controversy which follows the subject.
\newline
For an accurate meter, we were able to evaluate the mean shift of the pointer using (\ref{a5a}), calculate {\it independently} the mean value of the measured quantity in \Eq(\ref{a6}), and find a perfect agreement between the two. 
For an inaccurate (weak) meter, we can still evaluate the mean shift in (\ref{a7}), but do not know how to calculate the mean value of the projector in the presence of interference. One possible course of action 
is to use the similarity between Eqs.(\ref{a5a}) and (\ref{a7}), and  {\it define} its intrinsic mean value to be the complex valued quantity in the curly brackets in (\ref{a7}). Although this probability amplitude  already has a name, we can follow \cite{Ah1} in re-branding it as a 
"weak value of the projector $\Pii_1$ for a system pre- and post-selected in the states $|\psi\ra$ and $|\phi\ra$", 
$_\phi\la \Pii_1\ra_\psi$. The change is purely cosmetic, and our result still reads
 \begin{eqnarray}\label{a11}\nonumber
_\phi\la \Pii_1\ra_\psi \equiv \text{the (relative) probability amplitude to reach }
\\
\text{$|\phi\ra$ from $|\psi\ra$, via path $\{1\}$ in Fig.1b.}\q\q\n
 =\frac{\la \phi |\exp(-i\h T/2) |\Pii_1|\exp(-i\h T/2)|\psi\ra}{\la \phi |\exp(-i\h T)|\psi\ra}.
\end{eqnarray}
This definition is readily extended to an arbitrary operator $\hat{B}$, whose "weak value" $_\phi\la \hat{B}\ra_\psi=\frac{\la \phi |\exp(-i\h T/2) |\hat{B}|\exp(-i\h T/2)|\psi\ra}{\la \phi |\exp(-i\h T)|\psi\ra}$,
reduces to its original definition \cite{Ah1} for $\h=0$,
 \begin{eqnarray}\label{a11}
 _\phi\la \hat{B}\ra_\psi=\frac{\la \phi |\hat{B}|\psi\ra}{\la \phi |\psi\ra}, 
\end{eqnarray} 
and is the sum of all such amplitudes, weighted by the eigenvalues $B_i$.
The properties of probability amplitudes are well known, some of them have been discussed above, and there is nothing unusual 
about the weak values so far. 
\newline
 The main controversy stems from \cite{Ah1}, and is of its author's own making. The authors considered an inaccurate measurement of the $z$-component of a spin $1/2$, pre- and post-selecting it in two nearly orthogonal states.
 They found a mean pointer shift to be $100$, which is hardly surprising, given that the corresponding 
 relative probability amplitudes in \Eq(\ref{a9}) are large, since the transition is unlikely and $\sum_{i=1}^2 A_i^{\phi \gets \psi}$ is small. Yet in \cite{Ah1} this outcome is presented as an  "unusual" result of a "usual" measuring procedure.
 \newline
 The two quoted adjectives should, in fact, be interchanged. The measuring procedure is hardly a usual one, since in the chosen regime the meter ceases to destroy interference between measured alternatives, which according to Bohm \cite{Bohm}
 is likely to lead to "absurd results". This is readily seen from our analysis, but may be less clear in original approach used in \cite{Ah1}, where the authors chose to reduce the coupling to the pointer, instead of broadening its initial state. 
 The two methods are equivalent, since scaling the pointers position $f \to f/\gamma$ is equivalent 
 to multiplying $\h_{int}$ in \Eq(\ref{a3a}) by $\gamma$.
 \newline
The result is, however, what one would expect. Above we have shown that a weak measurement must be able to
produce all possible results, with mean shifts that are large, small, negative and positive. 
This is necessary, if we are to satisfy the uncertainty principle which forbids to look inside the union of two interfering alternatives \cite{Kast}. Indeed, if a group of experimenters decide to make accurate measurements of the $z$-component of a spin $1/2$, using all possible initial and final states, they will be able to agree that all readings are either $1/2$ or $-1/2$, and draw further conclusions
about the nature of the studied system. If the same experiment is repeated with inaccurate weak meters, the measured shifts will lie anywhere on the real axis. Making the experimenters evaluate accurately {\it mean} shifts for each choice of $|\psi\ra$ and $|\phi\ra$, would not help either. The mean shifts will also lie everywhere on the real axis, and the researches will be able to agree only on that  "anything is possible". 

In summary, we note that the uncertainty principle has elegantly frustrated our attempt to answer the "which way?" question in the presence of interference.
We started with a theoretical notion of a probability amplitude and employed a weak meter,
 hoping to gain further insight into 
what happens when the alternatives interfere. 
In the end, in this practical way, we
arrived at nothing more  than the very probability amplitude we started with. 
\newline
Identification of the weak values with probability amplitudes has, however, the advantage of explaining most of WV's controversial properties, using only the notions from the first chapter of Feynman's textbook \cite{FeynL}.
Firstly, the existence of "anomalous" weak values, lying outside the spectrum of the measured operator, is a rule, rather then an exception. They are just as common as the "normal" weak values, which occur when all amplitudes have the same sign, 
and coincide with the accurate mean values if only one of the amplitudes has a non-zero value.
From this it is clear that no "anomalous" mean values can be found in a classical theory, where all physical probabilities are non-negative.
Secondly, the authors of \cite{Ah1} have measured a difference between two large relative amplitudes of opposite signs, 
 where an accurate measurement would give the difference between two probabilities, and should not be surprised by the large result.
In the three-box case \cite{Ahbook}, involving three virtual paths with the amplitudes $A_1^{\phi \gets \psi}=A_2^{\phi \gets \psi}=-A_3^{\phi \gets \psi}=1$, simultaneous weak measurements of the projectors on the first and the second paths both yield values of $1$.
This does not mean that the particle is "in two places at the same time", but simply confirms the relation between the amplitudes, already known to us. A similar simple analysis can be applied to explain other "surprising" results, obtained within the weak measurement formalism. 
Finally, Vaidman's observation \cite{Vaid} that "The weak value shifts exist if measured or not" comes out as trivial, given that the WV's are nothing but the probability amplitudes, or their combinations. 
\newline
It has  not been our intention to belittle the technological effort invested in experimental realisations of "weak measurements" \cite{Exp1}, or their practical application in metrology \cite{Metro}. In our view, such efforts can only be helped by clarifying the status 
of the measured quantities within the framework of elementary quantum mechanics.
\acknowledgements
Support of
{MINECO and the European Regional Development Fund FEDER, through the grants
FIS2015-67161-P (MINECO/FEDER,UE) (D.S.),  MTM2016-76329-R (AEI / FEDER,
EU) (E.A.) and MINECO MTM2013-46553-C3-1-P (E.A.)
are gratefully acknowledged. E.A. thanks for support Basque Government - ELKARTEK Programme, grant KK-2016/0002. This research is also supported by the Basque Government through the BERC 2014-2017 program and by the Spanish Ministry of Economy and Competitiveness MINECO: BCAM Severo Ochoa accreditation SEV-2013-0323.} 
\newline



\begin{thebibliography}{10}
\bibitem{FOOT} A small modification of the experimental setup would allow to measure also the imaginary part of $\alpha_1^{\phi \gets \psi}$ \cite{Rev1}.
\bibitem{FOOT2} For $\h = 0$ we replace $|\psi\ra \to \exp(-i\h T/2|\psi\ra)$,  $|\phi\ra \to \exp(i\h T/2)|\phi\ra$, 
and proceed as before.
\end{thebibliography}
\end{document}